\def\aj{{AJ}}

\def\apj{{ApJ}}

\def\mnras{{MNRAS}}
\def\nat{{Nature}}

\def\aap{{A\&A}}
\def\apss{{Ap\&SS}}

\def\pasj{{PASJ}}
\def\araa{{ARA\&A}}

\newcommand{\Ho}{\mbox{$H_0$}}         
\newcommand{\neo}{\mbox{$n_{e  0}$}}   
\newcommand{\dTo}{\mbox{$\Delta T_0$}} 
\newcommand{\Sxo}{\mbox{$S_{x  0}$}}   
\newcommand{\Sx}{\mbox{$S_{x}$}}       
\newcommand{\Lamo}{\mbox{$\Lambda_{e \mbox{\tiny H} 0}$}}
\newcommand{\LameH}{\mbox{$\Lambda_{e \mbox{\tiny H}}$}}

\newcommand{\Tcmb}{\mbox{$T_{CMB}$}}     
\newcommand{\Da}{\mbox{$D_{\!\mbox{\tiny A}}$}}       
\newcommand{\kms}{\mbox{km s$^{-1}$}}
\newcommand{\ksM}{\mbox{km s$^{-1}$ Mpc$^{-1}$}}

\newcommand{\lsim}{\lesssim}

\newcommand{\Om}{\mbox{$\Omega_m$}}
\newcommand{\Ol}{\mbox{$\Omega_\Lambda$}}

\newcommand{\OmM}{\Om}
\newcommand{\OmL}{\Ol}

\newcommand{\kb}{\mbox{$k_{\mbox{\tiny B}}$}}     
\newcommand{\kB}{\kb}

\documentclass[cup5b]{caps}
\usepackage{natbib,psfig,ociwsymp2,amssymb,graphicx}
\HeadText{E. D. Reese}

\begin{document}

\pagenumbering{arabic}

\author[]{ERIK D.~REESE\\The University of California, Berkeley}

\chapter{Measuring the Hubble Constant \\ with the Sunyaev-Zel'dovich Effect}

\begin{abstract}

Combined with X-ray imaging and spectral data, observations of the 
Sunyaev-Zel'dovich effect (SZE) can be used to determine direct distances to
galaxy clusters.  These distances are independent of the extragalactic
distance ladder and do not rely on clusters being standard candles or
rulers.  Observations of the SZE have progressed from upper limits to
high signal-to-noise ratio detections and imaging of the SZE.  SZE/X-ray
determined distances to galaxy clusters are beginning to trace out the
theoretical angular-diameter distance relation.  The current ensemble
of 41 SZE/X-ray distances to galaxy clusters imply a Hubble constant
of $\Ho \approx 61 \pm 3 \pm 18$ \ksM, where the uncertainties are
statistical followed by systematic at 68\% confidence.  With a sample
of high-redshift galaxy clusters, SZE/X-ray distances can be used to
measure the geometry of the Universe.  

\end{abstract}

\section{Introduction}

Analysis of the Sunyaev-Zel'dovich effect (SZE) and X-ray data provides a
method of directly determining distances to galaxy clusters at any
redshift.  Clusters of galaxies contain hot ($\kB T_e \approx 10$ keV)
gas, known as the intracluster medium (ICM), trapped in their
potential wells.  Cosmic microwave background (CMB) photons passing
through a massive cluster interact with the energetic ICM electrons
with a probability of $\tau \approx 0.01$.  This inverse-Compton
scattering preferentially boosts the energy of a scattered CMB photon, 
causing a small ($\lsim 1$ mK) distortion in the CMB spectrum, known
as the Sunyaev-Zel'dovich effect \citep{sunyaev1970, sunyaev1972}.
The SZE is proportional to the pressure integrated along the line of
sight, $\Delta T \propto \int n_e T_e d\ell$.  X-ray emission from the
ICM has a different dependence on the density $S_{\mbox{\tiny X}}
\propto \int n_e^2 \LameH d\ell$, where \LameH\ is the X-ray cooling
function.  Taking advantage of the different density dependences and
with some assumptions about the geometry of the cluster, the distance
to the cluster may be determined.  SZE and X-ray determined distances
are independent of the extragalactic distance ladder and provide
distances to high-redshift galaxy clusters.  This method does not rely
on clusters being standard candles or rulers and relies only on
relatively simple properties of highly ionized plasma.

The promise of direct distances has been one of the primary
motivations for SZE observations.  Efforts over the first two decades
after the SZE was first proposed in 1970
\citep{sunyaev1970,sunyaev1972} yielded few reliable detections.  Over
the last decade, new detectors and observing techniques have allowed
high-quality detections and images of the effect for more than 60
clusters with redshifts as high as one.  SZE observations are routine
enough to build up samples of clusters and place constraints on
cosmological parameters.

The SZE offers a unique and powerful observational tool for
cosmology.  Distances to galaxy clusters yield a measurement of the
Hubble constant, $H_0$.  With a sample of high-redshift clusters, SZE
and X-ray distances can be used to determine the geometry of the
Universe.  In addition, the SZE has been used to measure gas fractions
in galaxy clusters \citep[e.g.,][]{myers1997, grego2001}, which can be
used to measure the matter density of the Universe, $\Omega_m$,
assuming the composition of clusters represents a fair sample of the
universal composition.  Upcoming deep, large-scale SZE surveys will
measure the evolution of the number density of galaxy clusters, which
is critically dependent on the underlying cosmology.  In principle, the
equation of state of the ``dark energy'' may be determined from the
evolution of the number density of clusters.

In this review, we first outline the properties of the SZE in
\S\ref{sec:sze} and provide a brief overview of the current state of
the observations in \S\ref{sec:obs_status}.  SZE/X-ray determined
distances are discussed in \S\ref{sec:cosmic_da}, briefly discussing
the current state of SZE/X-ray distances, sources of systematics, and
future potential. SZE surveys are briefly discussed in
\S\ref{sec:surveys} and a summary is given in
\S\ref{sec:summary}.  The physics of the SZE is covered in previous
reviews \citep{birkinshaw1999, rephaeli1995, sunyaev1980}, with
\citet{birkinshaw1999} and \citet{carlstrom2000} providing reviews of
the observations.  \citet{carlstrom2002} provide a review with focus
on cosmology from the SZE, with special attention to SZE surveys.

\section{The Sunyaev-Zel'dovich Effect}
\label{sec:sze}

\begin{figure*}[t]
\includegraphics[width=1.0\columnwidth,angle=0,clip]{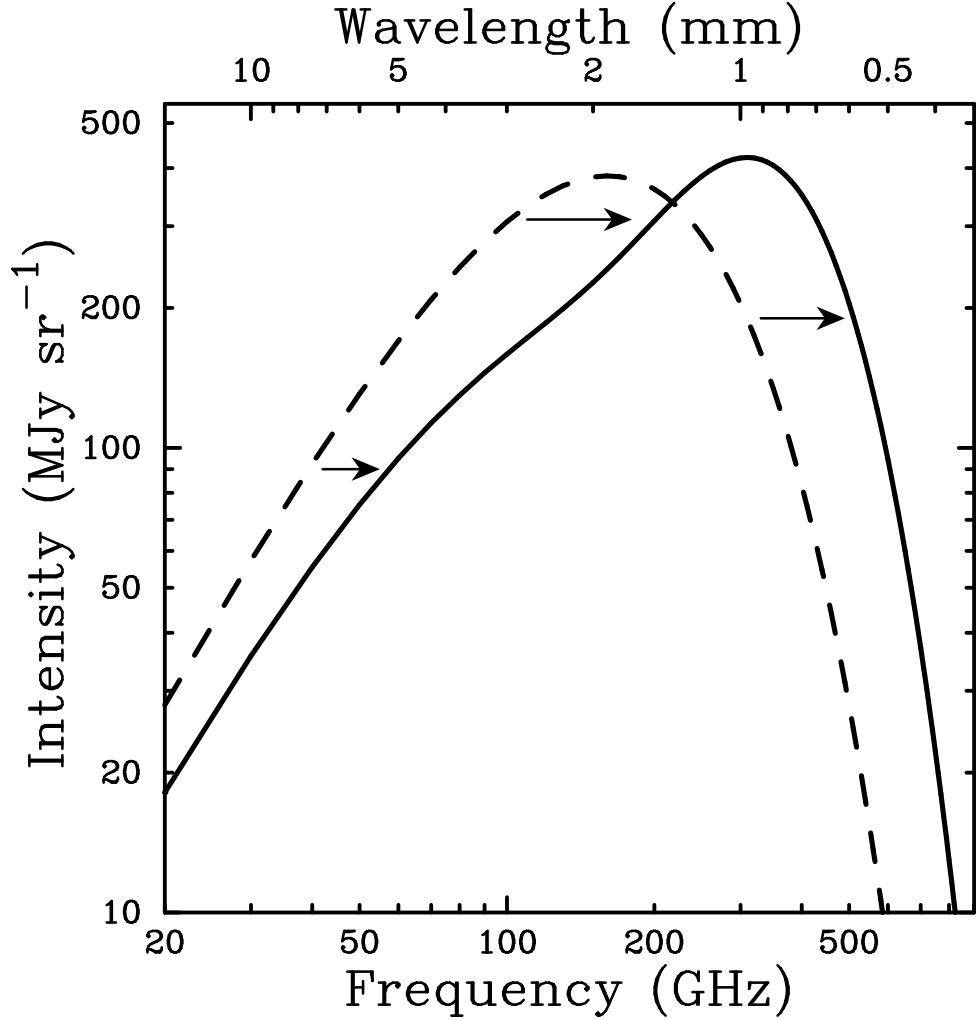}
\vskip 0pt \caption{
The CMB spectrum, undistorted (dashed line) and distorted by the SZE (solid 
line).  Following \cite{sunyaev1980}, to illustrate the effect, the SZE 
distortion shown is for a fictional cluster 1000 times more massive than a 
typical massive galaxy cluster.  The SZE causes a decrease in the CMB 
intensity at frequencies $\lsim 218$ GHz ($\sim 1.4$ mm) and an increase at 
higher frequencies.
\label{fig:sze_cmb}}
\end{figure*}

\begin{figure*}[t]
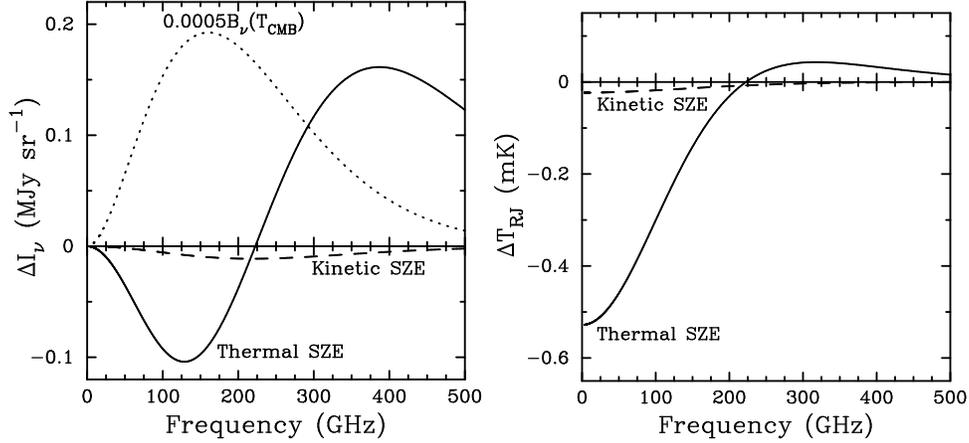

\includegraphics[width=0.49\columnwidth,angle=0,clip]{f2a.ps} \hfil
\includegraphics[width=0.49\columnwidth,angle=0,clip]{f2b.ps}
\vskip 0pt \caption{
  Spectral distortion of the CMB radiation due to the SZE.  The
  left panel shows the intensity, and the right panel shows the
  Rayleigh-Jeans brightness temperature.  The thick solid line is the
  thermal SZE, and the dashed line is the kinetic SZE.  For reference
  the 2.7 K thermal spectrum for the CMB intensity, scaled by 0.0005, is
  shown by the dotted line in the left panel.  The cluster properties
  used to calculate the spectra are an electron temperature of 10 keV,
  a Compton $y$-parameter of $10^{-4}$, and a peculiar velocity of 500 $\kms$.
\label{fig:sze_dI-dT}}
\end{figure*}

\subsection{Thermal Sunyaev-Zel'dovich Effect}
\label{subsec:thermal_sze}

The SZE is a small spectral distortion of
the CMB spectrum caused by the
scattering of the CMB photons off a distribution of high-energy
electrons. We only consider the SZE caused by the hot thermal
distribution of electrons provided by the ICM
of galaxy clusters.  CMB photons passing through the center of a
massive cluster have a $\tau_e \approx 0.01$ probability of
interacting with an energetic ICM electron.  The resulting inverse-Compton 
scattering preferentially boosts the energy of the CMB photon, 
causing a small ($\lsim 1$ mK) distortion in the CMB spectrum.  To 
illustrate the small effect,
Figure~\ref{fig:sze_cmb} shows the SZE spectral distortion for a
fictional cluster that is over 1000 times more massive than a typical
cluster.  The SZE appears as a decrease
in the intensity of the CMB at frequencies below $\lsim 218$ GHz and
as an increase at higher frequencies.

The SZE spectral distortion of the CMB, expressed as a
temperature change $\Delta T_{SZE}$ at dimensionless frequency $x
\equiv (h\nu)/(k_BT_{CMB})$, is given by
\begin{equation}
   \frac{\Delta T_{SZE}}{T_{CMB}} =  f(x) \ y  = f(x) \int
   n_e \frac{k_B T_e}{m_e c^2} \sigma_T \, d\ell, 
   \label{eq:deltaT1}\label{eq:y}
\end{equation}
where $y$ is the Compton $y$-parameter, which for an isothermal
cluster equals the optical depth times the fractional energy gain per
scattering. Here, $\sigma_T$ is the Thomson cross-section, $n_e$ is the
electron number density, $T_e$ is the electron temperature, $k_B$ is the
Boltzmann's constant, $m_e c^2$ is the electron rest-mass energy, and the
integration is along the line of sight.  The frequency dependence of
the SZE is
\begin{equation}
   f(x) = \left(x \frac{e^x+1}{e^x-1} -4\right)[1 + \delta_{SZE}(x,T_e)],
   \label{eq:fx}
\end{equation}
where $\delta_{SZE}(x,T_e)$ is the relativistic correction to the
frequency dependence.  Note that $f(x) \rightarrow -2$ in the
nonrelativistic and Rayleigh-Jeans (RJ) limits.  

It is worth noting that $\Delta T_{SZE} / \Tcmb$ is independent of
redshift, as shown in Equation~\ref{eq:deltaT1}. This unique feature of the
SZE makes it a potentially powerful tool for investigating the high-redshift 
Universe.

Expressed in units of specific intensity, common in millimeter
SZE observations, the thermal SZE is
\begin{equation}
\Delta I_{SZE} = g(x) I_0 y,
\label{eq:sze_intensity}
\end{equation}
where $I_0 = 2 (\kB\Tcmb)^3 / (h c)^2$ and the frequency dependence is
given by
\begin{equation}
g(x) = \frac{x^4 e^x}{(e^x-1)^2} \left ( x \frac{e^x + 1}{e^x - 1} - 4
\right ) \left [ 1 + \delta_{SZE}(x, T_e) \right ].
\label{eq:gx}
\end{equation}
$\Delta T_{SZE}$ and $\Delta I_{SZE}$ are simply related by the
derivative of the blackbody with respect to temperature, $\left |
dB_\nu / dT \right |$.

The spectral distortion of the CMB spectrum by the thermal SZE is
shown in Figure~\ref{fig:sze_dI-dT} (solid line) for a realistic
massive cluster ($y = 10^{-4}$), in units of intensity (left panel) and
RJ brightness temperature (right panel).  The RJ
brightness is shown because the sensitivity of a radio telescope is
calibrated in these units. It is defined simply by $I_\nu = (2 k_B
\nu^2/c^2) T_{RJ}$, where $I_\nu$ is the intensity at frequency $\nu$,
$k_B$ is Boltzmann's constant, and $c$ is the speed of light.  The CMB
blackbody spectrum, $B_\nu (\Tcmb)$, multiplied by 0.0005 (dotted
line), is also shown for comparison.  Note that the spectral signature
of the thermal effect is distinguished readily from a simple
temperature fluctuation of the CMB.  The kinetic SZE distortion is
shown by the dashed curve (\S\ref{sec:kinetic_sze}).  In the nonrelativistic 
regime, it is indistinguishable from a CMB temperature fluctuation.

The gas temperatures measured in massive galaxy clusters are around
$k_B T_e \approx 10$ keV \citep{mushotzky1997, allen1998} and are
measured to be as high as $\sim 17$ keV in the galaxy cluster 1E
$0657-56$ \citep{tucker1998}. At these temperatures, electron
velocities are becoming relativistic, and small corrections are
required for accurate interpretation of the SZE.  There has been
considerable theoretical work to include relativistic corrections to
the SZE 
\citep{wright1979, fabbri1981, rephaeli1995, rephaeli1997,
stebbins1997, challinor1998, itoh1998, nozawa1998, sazonov1998, sazonov1998b,
challinor1999, molnar1999, dolgov2001}.  
All of these
derivations agree for $\kb T_e \lsim 15$ keV, appropriate for galaxy
clusters.  For a massive cluster with $k_BT_e \approx 10$ keV
($k_BT_e/m_ec^2 \approx  0.02$), the relativistic corrections to the SZE
are of order a few percent in the RJ portion of the spectrum, but can
be substantial near the null of the thermal effect. Convenient
analytical approximations to fifth order in $k_BT_e/m_e c^2$ are
presented in \citet{itoh1998}.

\begin{figure*}[t]
\includegraphics[width=1.0\columnwidth,angle=0,clip]{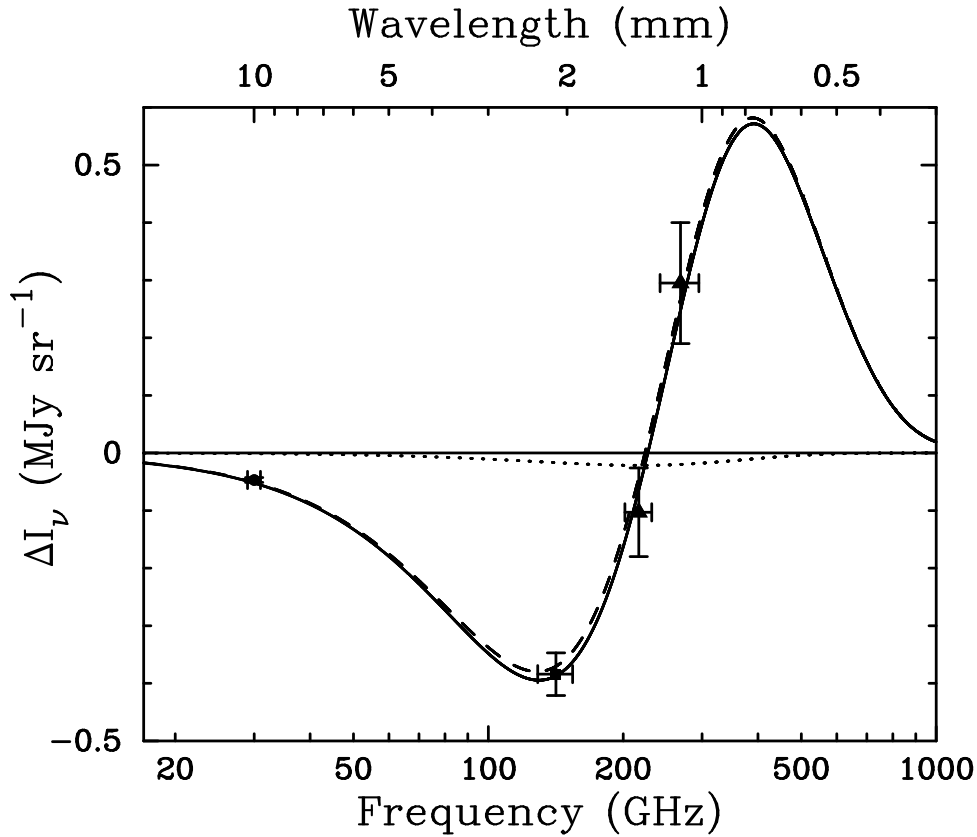}
\vskip 0pt \caption{
  The measured SZE spectrum of Abell 2163. The data point at
  30 GHz is from BIMA \citep{laroque2003b}, at 140~GHz is the
  weighted average of Diabolo and SuZIE measurements (filled square;
  Holzapfel et al. 1997a; D\'esert et al. 1998), and at 218~GHz and 270~GHz
  from SuZIE (filled triangles; Holzapfel et al. 1997a).
  Uncertainties are at 68\% confidence with the FWHM of the observing
  bands shown.  The best-fit thermal and kinetic SZE spectra are
  shown by the dashed line and the dotted lines, respectively, with
  the spectra of the combined effect shown by the solid line. The
  limits on the Compton $y$-parameter and the peculiar velocity are
  $y_0=3.71^{+0.36}_{-0.36}$$^{+0.33}_{-0.16}\times10^{-4}$ and
  $\upsilon_p=320^{+880}_{-740}$$^{+480}_{-440}$ km s$^{-1}$, respectively,
  with statistical followed by systematic uncertainties at 68\%
  confidence \citep{holzapfel1997b,laroque2003b}.
\label{fig:a2163_spec}}
\end{figure*}

The measured SZE spectrum of Abell 2163, spanning the decrement and
increment with data obtained from different telescopes and techniques,
is shown in Figure~\ref{fig:a2163_spec}
\citep{holzapfel1997b,desert1998,laroque2003b}.  Also plotted is the
best-fit model (solid) consisting of thermal (dashed) and kinetic
(dotted) SZE components.  The SZE spectrum is a good fit to the data,
demonstrating the consistency and robustness of modern SZE
measurements.

The most important features of the thermal SZE are: 
(1) it is a small spectral distortion of the CMB, of order $\sim 1$ mK, which 
is proportional to the cluster pressure integrated along the line of sight
(Eq.~\ref{eq:y}); 
(2) it is independent of redshift;  and
(3) it has a unique spectral signature with a decrease in the
CMB intensity at frequencies $\lsim 218$ GHz and an increase
at higher frequencies.

\subsection{Kinetic Sunyaev-Zel'dovich Effect}
\label{sec:kinetic_sze}

If the cluster is moving with respect to the CMB rest frame, there will
be an additional spectral distortion due to the Doppler effect of the
cluster bulk velocity on the scattered CMB photons.  If a component of
the cluster velocity, $\upsilon_{pec}$, is projected along the line of sight
to the cluster, then the Doppler effect will lead to an observed
distortion of the CMB spectrum, referred to as the kinetic SZE. In the
nonrelativistic limit, the spectral signature of the kinetic SZE is a
pure thermal distortion of magnitude
\begin{equation}
\frac{\Delta T_{SZE}}{T_{CMB}}  = -\tau_e \left ( \frac{\upsilon_{pec}}{c} 
\right ),
\label{eq:v_pec}
\end{equation}
where $\upsilon_{pec}$ is along the line of sight; that is, the emergent
spectrum is still described completely by a Planck spectrum, but at a
slightly different temperature, lower (higher) for positive (negative)
peculiar velocities \citep{sunyaev1972, phillips1995,
birkinshaw1999}.   Figure~\ref{fig:sze_dI-dT} illustrates the kinetic
SZE (dashed) for a typical galaxy cluster with a peculiar velocity of
500 $\kms$.   Figure~\ref{fig:a2163_spec} shows the SZE spectrum of
the galaxy cluster A2163 along with the best-fit model consisting of
thermal (dashed) and kinetic (dotted) SZE components.

Relativistic perturbations to the kinetic SZE are due to the Lorentz
boost to the electrons provided by the bulk velocity \citep{nozawa1998b,
sazonov1998}. The leading term is of order 
$(k_B T_e/m_e c^2)(\upsilon_{pec}/c)$,
 and for a 10~keV cluster moving at 1000 \kms\ the
effect is about an 8\% correction to the nonrelativistic term. The
$(k_B T_e/m_e c^2)^2(\upsilon_{pec}/c)$ term is only about 1\% of
the nonrelativistic kinetic SZE, and the
$(\upsilon_{pec}/c)^2$ term is only 0.2\%.

\section{Measurements of the SZE}
\label{sec:obs_status}

In the 20 years following the first papers by Sunyaev \& 
Zel'dovich (1970, 1972), there were few firm
detections of the SZE despite a considerable amount of effort
\citep[see][for a review of early experiments]{birkinshaw1999}. Over
the last several years, however, observations of the effect have
progressed from low signal-to-noise ratio detections and upper limits to
high-confidence detections and detailed images. In this section we
briefly review the current state of SZE observations.

The dramatic increase in the quality of the observations is due to
improvements both in low-noise detection systems and in observing
techniques, usually using specialized instrumentation to control
carefully the systematics that often prevent one from obtaining the
required sensitivity.  The sensitivity of a low-noise radio
receiver available 20 years ago should have easily allowed the
detection of the SZE toward a massive cluster. Most attempts, however,
failed due to uncontrolled systematics.  Now that the sensitivities of
detector systems have improved by factors of 3 to 10, it is clear that
the goal of all modern SZE instruments is the control of systematics.
Such systematics include, for example, the spatial and temporal
variations in the emission from the atmosphere and the surrounding
ground, as well as gain instabilities inherent to the detector system
used.

The observations must be conducted on the appropriate angular scales.
Galaxy clusters have a characteristic size scale of order a Mpc.
For a reasonable cosmology, a Mpc subtends an
arcminute or more at any redshift. Low-redshift clusters will subtend
a much larger angle; for example, the angular extent of the Coma
cluster ($z = 0.024$) is of order a degree (core radius $\sim 10'$; Herbig 
et al. 1995).  The detection of extended low-surface brightness
objects requires precise differential measurements made toward widely
separated directions on the sky. The large angular scale presents
challenges to control offsets due to differential ground pick-up and
atmospheric variations.

\subsection{Sources of Astronomical Contamination and Confusion}
\label{sec:confusion}

There are three main sources of astronomical contamination and confusion
that must be considered when observing the SZE: (1) CMB primary
anisotropies, (2) radio point sources, and (3) dust from the Galaxy and
external galaxies.  For distant clusters with angular extents of a few
arcminutes or less, the CMB anisotropy is expected \citep{hu1997} and
found to be damped considerably on these scales (\citealt{church1997,
subrahmanyan2000, dawson2001}; see also Holzapfel et al. 1997b and 
LaRoque et al. 2003 for CMB limits to SZE contamination).  For nearby
clusters, or for searches for distant clusters using beams larger than
a few arcminutes, the intrinsic CMB anisotropy must be considered.
The unique spectral behavior of the thermal SZE can be used to
separate it from the intrinsic CMB in these cases. Note, however, that
for such cases it will not be possible to separate the kinetic SZE
effects from the intrinsic CMB anisotropy without relying on the very
small spectral distortions of the kinetic SZE due to relativistic
effects.

Historically, the major source of contamination in the measurement of
the SZE has been radio point sources. It is obvious that emission from
point sources located along the line of the sight to the cluster could
fill in the SZE decrement, leading to an underestimate. Radio point
sources can also lead to overestimates of the SZE decrement --- for
example, point sources in the reference fields of single-dish
observations.  The radio point sources are variable and therefore must
be monitored. Radio emission from the cluster member galaxies, from
the central dominant (cD) galaxy in particular, is often the largest source of
radio point source contamination, at least at high radio frequencies
\citep{cooray1998a, laroque2003b}.

At frequencies near the null of the thermal SZE and higher, dust
emission from extragalactic sources as well as dust emission from our
own Galaxy must be considered.  At the angular scales and frequencies
of interest for most SZE observations, contamination from diffuse
Galactic dust emission will not usually be significant and is easily
compensated.  Consider instead the dusty extragalactic sources such as
those that have been found toward massive galaxy clusters with the
SCUBA bolometer array \citep{smail1997}. Spectral indices for these
sources are estimated to be $\sim 1.5-2.5$
\citep{fischer1993,blain1998}.  Sources with 350~GHz (850~$\mu$m)
flux densities greater than 8~mJy are common, and all clusters surveyed had
multiple sources with flux densities greater than 5~mJy.  This translates into
an uncertainty in the peculiar velocity for a galaxy cluster of
roughly 1000 $\kms$ (see \citealt{carlstrom2002} for details).

As with SZE observations at radio frequencies, the analyses of
high-frequency observations must also consider the effects of point
sources and require either high dynamic angular range, large spectral
coverage, or both to separate the point source emission from the SZE.

\begin{figure*}[t]
\includegraphics[width=1.0\columnwidth,angle=0,clip]{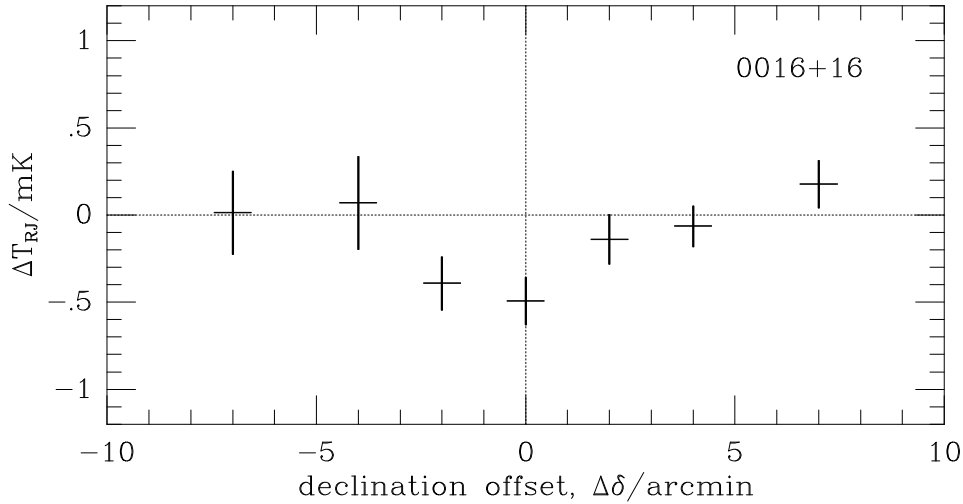}
\vskip 0pt \caption{
Measurement of the SZE profile across the galaxy cluster Cl~$0016+16$ obtained 
with the OVRO 40~m telescope \citep{hughes1998}.  The observed profile 
provided confidence in the reliability of SZE detections.
\label{fig:ovro40m_cl0016}}
\end{figure*}

\subsection{Single-dish Observations}
\label{sec:single_dish}

The first measurements of the SZE were made with single-dish radio
telescopes at centimeter wavelengths.  Advances in detector technology
made the measurements possible, although early observations appear to
have been plagued by systematic errors that led to irreproducible and
inconsistent results.  Eventually, successful detections using beam-switching 
techniques were obtained.  During this period, the
pioneering work of Birkinshaw and collaborators with the OVRO 40~m
telescope stands out for its production of results that served to
build confidence in the technique \citep{birkinshaw1978a,
birkinshaw1978b,birkinshaw1991}.  Figure~\ref{fig:ovro40m_cl0016}
shows the OVRO 40~m scan through the galaxy cluster Cl $0016+16$. More
recently, leading and trailing beam-switching techniques have been
used successfully with the OVRO 5~m telescope at 32 GHz to produce
reliable detections of the SZE in several intermediate-redshift
clusters \citep{herbig1995, myers1997, mason2001}. The SEST 15~m
and IRAM 30~m telescopes have been used with bolometric detectors
at 140 GHz and chopping mirrors to make significant detections of the
SZE in several clusters \citep{andreani1996, andreani1999,
desert1998, pointecouteau1999, pointecouteau2001}.  The Nobeyama 45~m
telescope has also been been used at 21, 43, and 150~GHz to
detect and map the SZE \citep{komatsu2001, komatsu1999}.

The Sunyaev-Zel'dovich Infrared Experiment (SuZIE) uses its six-element 
140 GHz bolometer array to observe in a drift-scanning mode,
where the telescope is fixed and the rotation of the Earth moves the
beams across the sky.  Using this drift-scanning technique, the SuZIE
experiment has produced high signal-to-noise ratio strip maps of the SZE
emission in several clusters \citep{holzapfel1997b,mauskopf2000}.

Because of the high sensitivity of bolometric detectors at millimeter
wavelengths, single-dish experiments are ideally suited for the
measurement of the SZE spectrum.  By observing at several millimeter
frequencies, these instruments should be able to separate the thermal
and kinetic SZEs from atmospheric fluctuations and sources of
astrophysical confusion.  

The measured SZE spectrum of Abell 2163, spanning the decrement and
increment with data obtained from different telescopes and techniques,
is shown in Figure~\ref{fig:a2163_spec}
\citep{holzapfel1997b,desert1998,laroque2003b}.  The SZE spectrum is a
good fit to the data, demonstrating the consistency and robustness of
modern SZE measurements.

Single-dish observations of the SZE are just beginning to reach their
potential, and the future is very promising.  The development of large-format, 
millimeter-wavelength bolometer arrays will increase the
mapping speed of current SZE experiments by orders of magnitude.  To
the extent that atmospheric fluctuations are common across a
bolometric array, it will be possible to realize the intrinsic
sensitivity of the detectors.  Operating from high astronomical sites
with stable atmospheres and exceptionally low precipitable water
vapor, future large-format bolometer arrays have the potential to
produce high signal-to-noise ratio SZE images and search for distant SZE
clusters with unprecedented speed.

\subsection{Interferometric Observations}
\label{sec:interferometer}

The stability and spatial filtering inherent to interferometry has
been exploited to make high-quality images of the SZE.  The stability
of an interferometer is due to its ability to perform simultaneous
differential sky measurements over well-defined spatial frequencies.
The spatial filtering of an interferometer also allows the emission
from radio point sources to be separated from the SZE emission. 

There are several other features that allow
an interferometer to achieve extremely low systematics. For example,
only signals that correlate between array elements will lead to
detected signal. For most interferometers, this means that the bulk of
the sky noise for each element will not lead to signal. Amplifier gain
instabilities for an interferometer will not lead to large offsets or
false detections, although, if severe, they may lead to somewhat noisy
signal amplitude. To remove the effects of offsets or drifts in the
electronics, as well as the correlation of spurious (noncelestial)
sources of noise, the phase of the signal received at each telescope
is modulated, and then the proper demodulation is
applied to the output of the correlator.

The spatial filtering of an interferometer also allows the emission
from radio point sources to be separated from the SZE emission. This
is possible because at high angular resolution ($\lsim 10''$) the
SZE contributes very little flux.  This allows one to use long
baselines, which give high angular resolution, to detect and
monitor the flux of radio point sources, while using short baselines to
measure the SZE.  Nearly simultaneous monitoring of the point sources
is important, as they are often time variable.  The signal from the
point sources is then easily removed, to the limit of the dynamic
range of the instrument, from the short-baseline data, which are
sensitive also to the SZE.

For the reasons given above, interferometers offer an ideal way to
achieve high brightness sensitivity for extended low-surface
brightness emission, at least at radio wavelengths.  Most
interferometers, however, were not designed for imaging low-surface
brightness sources.  Interferometers have been built traditionally to
obtain high angular resolution and thus have employed large individual
elements for maximum sensitivity to small-scale emission.  As a result,
special-purpose interferometric systems have been built for imaging
the SZE \citep{jones1993,carlstrom1996,padin2001}.  All of them have
taken advantage of low-noise HEMT amplifiers \citep{pospieszalski1995}
to achieve high sensitivity.

The first interferometric detection \citep{jones1993} of the SZE was
obtained with the Ryle Telescope (RT).  The RT was built from the 5
Kilometer Array, consisting of eight 13 m telescopes located in
Cambridge, England, operating at 15~GHz with East-West
configurations. Five of the telescopes can be used in a compact E-W
configuration for imaging of the SZE \citep{jones1993, grainge1993,
grainge1996, grainge2002, grainge2002b,
grainger2002, jones2001,saunders1999}.

\begin{figure*}[t]
\includegraphics[width=1.0\columnwidth,angle=0,clip]{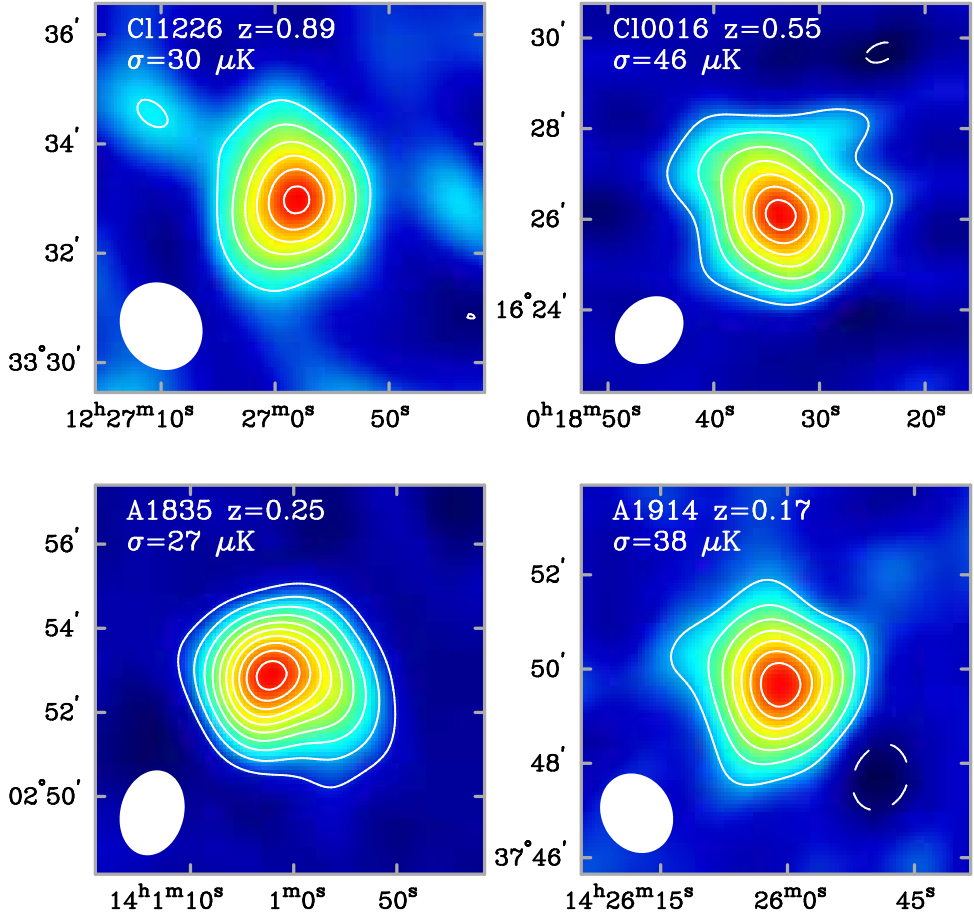}
\vskip 0pt \caption{
  Deconvolved interferometric SZE images for four galaxy
  clusters over a large redshift range ($0.17 \leq z \leq 0.89$).  The
  contours are multiples of 2$\sigma$, and negative contours are shown
  as solid lines.  The FWHM ellipse of the synthesized beam (PSF) is
  shown in the lower-left corner of each panel. The rms, $\sigma$,
  appears in the top of each panel. Radio point sources were removed
  from three of the images shown.  The interferometer was able to separate
  the point source emission from the SZE by using the high-resolution
  data obtained with long baselines. All of the clusters shown have
  similarly high X-ray luminosities, and, as can be seen, the strength of
  the SZE signals are similar despite the factor of 5 in redshift,
  illustrating the independence of the SZE on redshift.
\label{fig:szpanel12}}
\end{figure*}

The OVRO and BIMA SZE imaging project uses 30~GHz (1 cm) low-noise
receivers mounted on the OVRO\footnote{An array of six 10.4 m
telescopes located in the Owens Valley, CA, operated by Caltech.}
and BIMA\footnote{An array of 10 6.1 m mm-wave telescopes located at
Hat Creek, CA, operated by the
Berkeley-Illinois-Maryland-Association.} mm-wave arrays in
California. They have produced SZE images toward 60 clusters to date
\citep{carlstrom1996, carlstrom2000, grego2000, grego2001, patel2000, 
reese2000, reese2002, joy2001, laroque2003b}.  A sample of their SZE
images is shown in Figure~\ref{fig:szpanel12}.  All contours are
multiples of 2 $\sigma$ of each image, and the full-width at half
maximum (FWHM) of the synthesized beam (PSF for this deconvolution) is shown
in the lower left-hand corner of each image.
Figure~\ref{fig:szpanel12} also clearly demonstrates the independence
of the SZE on redshift. All of the clusters shown have similarly high
X-ray luminosities, and, as can be seen, the strength of the SZE
signals are similar despite the factor of 5 in redshift.  The OVRO
and BIMA arrays support two-dimensional configurations of the
telescopes, including extremely short baselines, allowing good
synthesized beams for imaging the SZE of clusters at declinations
greater than $\sim -15$~degrees.

The RT, OVRO, and BIMA SZE observations are insensitive to the angular
scales required to image low-redshift ($z \ll 0.1$) clusters.  Recently,
however, the Cosmic Background Imager (CBI; Padin et al. 2001) has been
used to image the SZE in a few nearby clusters
\citep{udomprasert2000}.  The CBI is composed of 13 0.9~m
telescopes mounted on a common platform, with baselines spanning 1 m to
6 m. Operating in 10 1~GHz channels spanning 26 -- 36 GHz, it is
sensitive to angular scales spanning 3$'$ to 20$'$.  The large field
of view of the CBI, 0.75~ degrees FWHM, makes it susceptible to
correlated contamination from terrestrial sources (i.e., ground
emission). To compensate, they have adopted the same observing strategy
as for single-dish observations (\S\ref{sec:single_dish}), by
subtracting from the cluster data, data from leading and trailings fields
offset by $\pm$12.5 minutes in Right Ascension from the cluster.

Interferometric observations of the SZE, as for single-dish
observations, are just beginning to demonstrate their
potential. Upcoming instruments will be over an order of magnitude
more sensitive.  This next generation of interferometric SZE
instruments will conduct deep SZE surveys covering tens, and possibly
hundreds, of square degrees. While not as fast as planned large-format
bolometric arrays, the interferometers will be able to survey deeper
and provide more detailed imaging.  In particular, the high resolution
and deep imaging provided by future heterogeneous arrays will provide
a valuable tool for investigating cluster structure and its evolution.
Such studies are necessary before the full potential of large SZE
surveys for cosmology can be realized.

\section{The Cosmic Distance Scale from SZE/X-ray Distances}
\label{sec:cosmic_da}

Several years after the SZE was first proposed \citep{sunyaev1970,
sunyaev1972} it was recognized that the distance to a cluster could be
determined with a measure of its SZE and X-ray emission
\citep{cavaliere1977, boynton1978, cavaliere1978, gunn1978, silk1978, 
birkinshaw1979}.  The distance is determined by exploiting the
different density dependences of the SZE and X-ray emissions.  The
SZE is proportional to the first power of the density; $\Delta T_{SZE}
\sim \int d\ell n_e T_e$, where $n_e$ is the electron density, $T_e$
is the electron temperature, and $d\ell$ is along the line-of-sight.
The distance dependence is made explicit with the substitution $d\ell =
\Da d\zeta$, where $\Da$ is the angular-diameter distance of the
cluster.

\begin{figure*}[t]
\includegraphics[width=1.0\columnwidth,angle=0,clip]{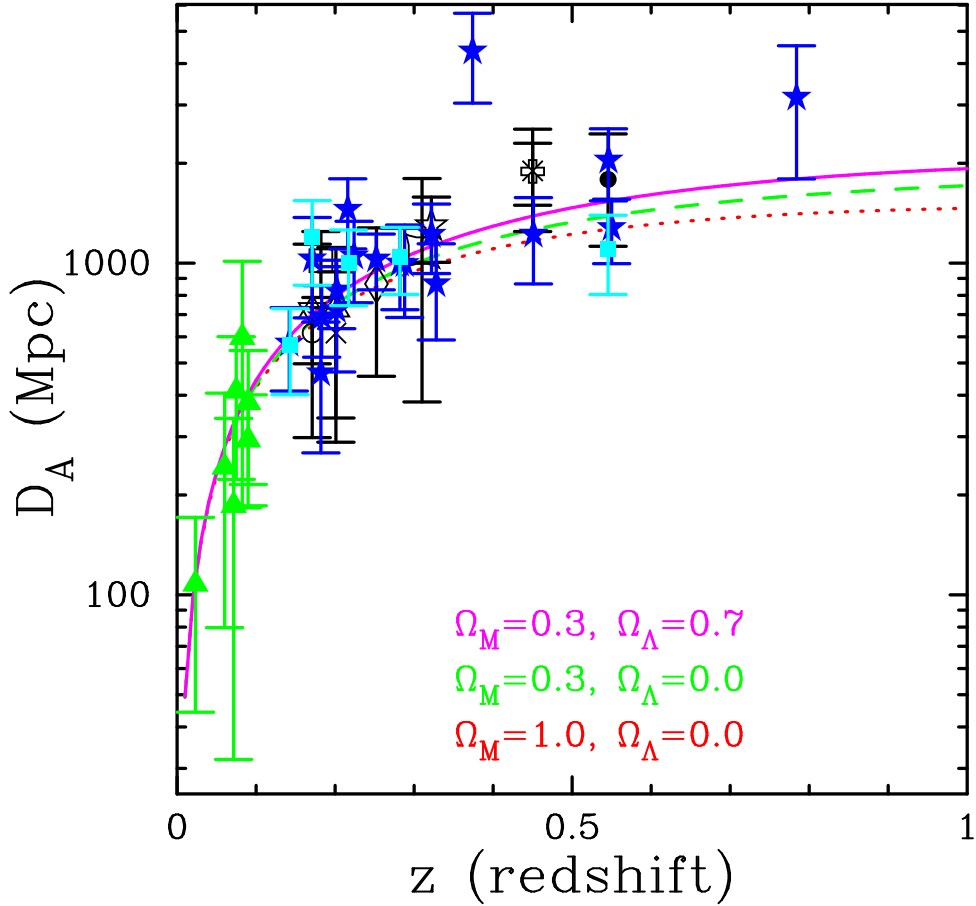} 
\vskip 0pt \caption{ SZE-determined distances versus redshift. The
theoretical angular diameter distance relation is plotted for three
different cosmologies, assuming $\Ho = 60$ \ksM: $\Lambda$ ---
$\OmM=0.3$, $\OmL=0.7$ (solid line), open --- $\OmM=0.3$ (dashed), and
flat --- $\OmM=1$ (dot-dashed).  The clusters are beginning to trace
out the angular-diameter distance relation.  Three samples are
highlighted: seven nearby clusters observed with the OVRO 5~m (green
solid triangles; Myers et al. 1997; Mason et al. 2001); five clusters
from Ryle (cyan solid squares; Grainge et al. 2002b; Jones et
al. 2003; Saunders et al. 2003); and 18 clusters from the OVRO/BIMA
SZE imaging project (blue solid stars; Reese et al. 2000, 2002).
Additional references: \citealt{birkinshaw1991, birkinshaw1994,
holzapfel1997, hughes1998, lamarre1998, tsuboi1998, andreani1999,
komatsu1999, mauskopf2000, patel2000, pointecouteau2001}.
\label{fig:da}}
\end{figure*}

The X-ray emission is proportional to the second power of the density;
$\Sx \sim \int d\ell n_e^2 \LameH$, where \LameH\ is the X-ray cooling
function.  The angular-diameter distance is solved for by eliminating
the electron density,\footnote{Similarly, one could eliminate \Da\ in
favor of the central density, \neo.} yielding
\begin{equation}
\Da \propto \frac{(\dTo)^2 \Lamo}{\Sxo T_{e 0}^2} \frac{1}{\theta_c},
	\label{eq:Dadepend}
\end{equation}
where these quantities have been evaluated along the line of sight
through the center of the cluster (subscript 0) and $\theta_c$ refers
to a characteristic scale of the cluster along the line of sight,
whose exact meaning depends on the density model adopted.  Only the
characteristic scale of the cluster in the plane of the sky is
measured, so one must relate the characteristic scales along the line of sight 
and in the plane of the sky.  For detailed treatments of this calculation,
see \citet{birkinshaw1991} and \citet{reese2000, reese2002}.  Combined
with the redshift of the cluster and the geometry of the Universe, one
may determine the Hubble parameter, with the inverse dependences on
the observables as that of $\Da$.  With a sample of galaxy clusters,
one fits the cluster distances versus redshift to the theoretical
angular-diameter distance relation, with the Hubble constant as the
normalization (see, e.g., Fig.~\ref{fig:da}).

\subsection{Current Status of SZE/X-ray Distances}
\label{sec:Da_now}

To date, there are 41 distance determinations to 26 different galaxy
clusters from analysis of SZE and X-ray observations.  All of these
SZE/X-ray distances use {\it ROSAT} X-ray data and model the cluster
gas as a spherical isothermal $\beta$ model \citep{cavaliere1976,
cavaliere1978}.  The {\it ROSAT} data do not warrant a more
sophisticated treatment.  In Figure~\ref{fig:da} we show all SZE-determined 
distances from high signal-to-noise ratio SZE experiments.  The
uncertainties shown are statistical at 68\% confidence.  The
theoretical angular-diameter distance relation is shown for three
cosmologies assuming $H_0 = 60$ $\ksM$.

There are currently three homogeneously analyzed samples of clusters
with SZE distances: (1) a sample of seven nearby ($z<0.1$) galaxy
clusters observed with the OVRO 5~m telescope (green solid triangles)
that finds $H_0 = 66
^{+14}_{-11}\, ^{+15}_{-15}$ \citep{myers1997, mason2001}, (2) a
sample of five intermediate-redshift ($0.14<z<0.3$) clusters from the
RT interferometer that finds $H_0 = 65 ^{+8}_{-7}\, ^{+15}_{-15}$
(cyan solid squares) \citep{jones2001}, and (3) a sample of 18
clusters with $0.14 < z < 0.83$ from interferometric observations by
the OVRO and BIMA SZE imaging project, which infers $H_0 = 60
^{+4}_{-4}\, ^{+18}_{-13}$ (blue solid stars) \citep{reese2002}.
The above Hubble constants assume a $\Om = 0.3$, $\Ol = 0.7$
cosmology, and the uncertainties are statistical followed by
systematic at 68\% confidence.  The treatment of uncertainties varies
among the three cluster samples.

A fit to the ensemble of 41 SZE-determined distances yields $\Ho
\approx 61 \pm 3 \pm 18$ km s$^{-1}$ Mpc$^{-1}$ for an $\Om = 0.3$,
$\Ol = 0.7$ cosmology, where the uncertainties are statistical
followed by systematic at 68\% confidence.  Since many of the clusters
are at high redshift, the best-fit Hubble constant will depend on the
cosmology adopted; the best-fit Hubble constant shifts to 57 km
s$^{-1}$ Mpc$^{-1}$ for an open $\Om = 0.3$ universe, and to 54 km
s$^{-1}$ Mpc$^{-1}$ for a flat $\Om = 1$ geometry.  The systematic
uncertainty, discussed below, clearly dominates.  The systematic
uncertainty is approximate because it is complicated by shared
systematics between some distance determinations.  For example,
including multiple distance determinations to a single cluster overstresses 
any effects of asphericity and orientation from that galaxy
cluster.

Statistical uncertainty includes contributions from the ICM shape
parameters, the electron temperature, point sources in the field, and
the cooling functions, which depends on $T_e$, metallicity, and the
column density.  The largest sources of statistical uncertainty are
the ICM shape parameters (from model fitting) and the X-ray determined
electron temperature ($\Da\propto T_e^{-2}$).  Uncertainty from model
fitting is roughly 20\% in the distance and is dominated by the
uncertainty in the central decrement.  The contribution from $T_e$ on
the distances varies greatly from $\sim 5$\% to $\sim 30$\%, with
10\%--20\% being typical values.  As expected, nearby cluster
temperatures are more precisely determined than those of distant
galaxy clusters.

\begin{table}
\caption{$H_0$ Systematic Uncertainty Budget \ \ \  \ \  \ \  \ \
\label{tab:H0_sys}}
    \begin{tabular}{lc}
     \hline \hline
     {Systematic} & {Effect (\%)} \\
     \hline
     SZE calibration            & $\pm 8$\\
     X-ray calibration          & $\pm 10$\\
     $N_H$                      & $\pm 5$\\
     Asphericity$^{*}$          & $\pm 5$\\
     Isothermality              & $\pm 10$\\
     Clumping                   & $-   20$\\
     Undetected radio sources   & $\pm 12$\\
     Kinetic SZE$^{*}$          & $\pm 2$\\
     Primary CMB$^{*}$          & $<\pm 1$\\
     Radio halos                & $-   3$\\
     Primary beam               & $\pm 3$\\
     $\ \ $Total                & $^{+22}_{-30}$\\
     \hline \hline
    \end{tabular}
  \label{sample-table}

\footnotesize{
$^{*}$Includes $1/\sqrt{18}$ factor for the 18 cluster sample.}
 \end{table}

\subsection{Sources of Possible Systematic Uncertainty}
\label{sec:Da_sys}

The absolute calibration of both the SZE and X-ray observations
directly affects the distance determinations.  In addition to the
absolute calibration uncertainty from the observations, there are
possible sources of systematic uncertainty that depend on the physical
state of the ICM and other sources that can contaminate the cluster
SZE emission.  Table~\ref{tab:H0_sys} summarizes the systematic
uncertainties in the Hubble constant determined from 30 GHz
interferometric SZE observations of a sample of 18 clusters
\citep{reese2002}, but are typical of most SZE experiments.  The
entries marked with asterisks are expected to average out for a large
sample of clusters and include a $1/\sqrt{18}$ factor reflecting the
18 clusters used in this work.  For detailed discussions of
systematics see \citet{birkinshaw1999} and \citet{reese2000, reese2002}.

\subsubsection{Cluster Structure}

Most clusters do not appear circular in radio, X-ray, or optical wavelengths.
Under the assumption of axisymmetric clusters, the combined effect of
cluster asphericity and its orientation on the sky conspires to
introduce a roughly $\pm 20$\% random uncertainty in $H_0$ determined
from one galaxy cluster \citep{hughes1998}.  When one considers a
large, unbiased sample of clusters, with random orientations, the
errors due to imposing a spherical model are expected to cancel,
resulting in a precise determination of $H_0$.  Numerical simulations
using triaxial $\beta$ models support this assumption
\citep{sulkanen1999}.

Departures from isothermality in the cluster atmosphere may result in
a large error in the distance determination from an isothermal
analysis.  The {\it ROSAT} band is fairly insensitive to temperature
variations, showing a $\sim 10$\% change in the PSPC count rate for a
factor of 2 change in temperature for $T_e > 1.5$ keV gas
\citep{mohr1999a}.  A mixture of simulations and studies of nearby
clusters suggests a 10\% effect on the Hubble parameter due to
departures from isothermality \citep[e.g.,][]{inagaki1995, roettiger1997}.

Clumping of the intracluster gas is a potentially
serious source of systematic error in the determination of the Hubble
constant.  Unresolved clumps in an isothermal intracluster plasma will
enhance the X-ray emission by a factor $C^2$, where
\begin{equation}
C \equiv \frac{\left < n_e^2 \right >^{1/2}}{\left < n_e \right >}.
\label{eq:C_def}
\end{equation}
If significant substructure exists in galaxy clusters, the cluster
generates more X-ray emission than expected from a uniform ICM,
leading to an underestimate of the angular-diameter distance ($\Da
\propto \Sxo^{-1}$) and therefore an overestimate of the Hubble
parameter by a factor $C^2$.  Unlike orientation bias, which
averages down for a large sample of clusters, clumping must be
measured in each cluster or estimated for an average cluster.  There
is currently no observational evidence of significant clumping in
galaxy clusters.  If clumping were significant and had large variations
from cluster to cluster, we might expect larger scatter than is seen
in the Hubble diagrams from SZE and X-ray distances
(Fig.~\ref{fig:da}).  In addition, the agreement between SZE 
\citep[e.g.,][]{grego2001} and X-ray \citep[e.g.,][]{mohr1999a}
determined gas fractions from galaxy clusters also suggests that
clumping is not a large effect.

\subsubsection{Possible SZE Contaminants}

Undetected point sources near the cluster mask the central decrement,
causing an underestimate in the magnitude of the decrement and
therefore an underestimate of the angular diameter distance.  Point
sources in reference fields and for interferometers, the complicated
synthesized beam shapes, may cause overestimates of the angular
diameter distance.  Massize clusters typically have central dominant
(cD) galaxies, which are often radio bright.  Therefore, it is likely
that there is a radio point source near the center of each cluster.
Typical radio sources have falling spectra, roughly $\alpha \approx 1$,
where $S_\nu \propto \nu^{-\alpha}$.  At 30 GHz, possible undetected
point sources just below the detection threshold of the observations
introduce a $\sim 10$\% uncertainty.

Cluster peculiar velocities with respect to the CMB introduce an
additional CMB spectral distortion known as the kinetic SZE (see
\S\ref{sec:kinetic_sze}).  For a single isothermal cluster, the ratio
of the kinetic SZE to the thermal SZE is
\begin{equation}
\left | \frac{\Delta T_{kinetic}}{\Delta T_{thermal}} \right | = 0.2
  \frac{1}{|f(x)|} 
  \left (\frac{\upsilon_{pec}}{1000 \mbox{ km s$^{-1}$}} \right )
  \left (\frac{10 \mbox{ keV}}{T_e} \right ),
\label{eq:kin_therm_frac}
\end{equation}
where $\upsilon_{pec}$ is the peculiar velocity along the line of sight. At
low frequencies (Rayleigh-Jeans regime), $f(x)\approx -2$ and the kinetic SZE
is $\sim 10$\% of the thermal SZE.  Recent observational evidence
suggests a typical one-dimensional rms peculiar velocity of $\sim 300$
$\kms$ \citep{watkins1997}, and recent simulations found similar
results \citep{colberg2000}.  In general, the kinetic effect is $\lsim 10$\%
that of the thermal SZE, except near the thermal null at $\sim 218$ GHz
where the kinetic SZE dominates (see Fig.~\ref{fig:sze_dI-dT}).
Cluster peculiar velocities are randomly distributed, so when averaged
over an ensemble of clusters, the effect from peculiar velocities
should cancel.

CMB primary anisotropies have the same spectral signature as the
kinetic SZE.  The effects of primary anisotropies on cluster distances depend
strongly on the beam size of the SZE observations and the typical
angular scale of the clusters being observed (nearby versus distant
clusters); the CMB effects on the inferred Hubble
constant should average out over an ensemble of clusters.  Recent BIMA
observations provide limits on primary anisotropies on scales of a few 
arcminutes \citep{holzapfel2000, dawson2001}.  On these scales, CMB
primary anisotropies are an unimportant ($\lsim 1$\%) source of
uncertainty.  For nearby clusters, or for searches for distant
clusters using beams larger than a few arcminutes, the intrinsic CMB
anisotropy must be considered.  The unique spectral signature of the
thermal SZE can be used to separate it from primary CMB anisotropy.
However, it will not be possible to separate primary CMB anisotropies
from the kinetic SZE without relying on the very small spectral
distortions of the kinetic SZE due to relativistic effects.

The SZE may be masked by large-scale diffuse nonthermal radio emission
in clusters of galaxies, known as radio halos.  If present, radio halos
are located at the cluster centers, have sizes typical of galaxy
clusters, and have a steep radio spectrum ($\alpha \approx  1 - 3$; 
Hanish 1982; Moffet \& Birkinshaw 1989; Giovannini et al. 1999; Kempner \& 
Sarazin 2001).  Because
halos are rare, little is known about their nature and origin, but
they are thought to be produced by synchrotron emission from an
accelerated or reaccelerated population of relativistic electrons
\citep[e.g.,][]{jaffe1977, dennison1980, roland1981,
schlickeiser1987}.  Conservative and simplistic modeling of the
possible effects of these halos implies a $\sim 3$\% overestimate on
the inferred Hubble parameter from radio halos \citep{reese2002}.
\citet{reese2002} show Very Large Array NVSS contours overlaid on 30 GHz
interferometric SZE images, suggesting that radio halos have little
impact on SZE observations and therefore on SZE/X-ray distances.

Imprecisely measured beam shapes affect the inferred central
decrements and therefore affect the Hubble constant.  For
interferometric observations, the primary beam is determined from
holography measurements.  Conservative and simple modeling of the
effects of the primary beam suggests that the effect on the Hubble
constant is a few percent ($\lsim 3$\%) at most.  In theory, this is a
controllable systematic with detailed measurements of beam shape and 
is currently swamped by larger potential sources of systematic
uncertainty.

\subsection{Future of SZE/X-ray Distances}
\label{sec:Da_future}

The prospects for improving both the statistical and systematic
uncertainties in the SZE distances in the near future are promising.
Note, from Equation~\ref{eq:Dadepend}, that the error budget in the
distance determination is sensitive to the absolute calibration of the
X-ray and SZE observations.  Currently, the best absolute calibration
of SZE observations is $\sim 2.5$\% at 68\% confidence, based on
observations of the brightness of the planets Mars and Jupiter.
Efforts are now underway to reduce this uncertainty to the 1\% level
(2\% in \Ho).  Uncertainty in the X-ray intensity scale also adds
another shared systematic. The accuracy of the {\it ROSAT} X-ray intensity
scale is debated, but a reasonable estimate is believed to be $\sim
10$\%. It is hoped that the calibration of the {\it Chandra}\ and 
{\it XMM-Newton}\ X-ray telescopes will greatly reduce this uncertainty.

Possible sources of systematic uncertainty are summarized in
Table~\ref{tab:H0_sys}.  These values come from 30 GHz interferometric
SZE observations and {\it ROSAT} data for a sample of 18 galaxy clusters
\citep{reese2002}, but are typical of most SZE distance determinations.
The largest systematic uncertainties are due to departures from
isothermality, the possibility of clumping, and possible point source
contamination of the SZE observations \citep[for detailed discussion
of systematics, see, e.g.,][]{birkinshaw1999, reese2000, reese2002}.
{\it Chandra}\ and {\it XMM-Newton}\ are already providing temperature 
profiles of galaxy clusters \citep[e.g.,][]{markevitch2000, nevalainen2000b,
tamura2001}.  The unprecedented angular resolution of {\it Chandra}\ will
provide insight into possible small-scale structures in clusters.  In
addition, multiwavelength studies by existing radio observatories, for example
the Very Large Array, can shed light on the residual point
source contamination of the radio wavelength SZE measurements.
Therefore, though currently at the 30\% level, many of the systematics
can and will be addressed through both existing X-ray and radio
observatories and larger samples of galaxy clusters provided from SZE
surveys.

\begin{figure*}[t]
\includegraphics[width=1.0\columnwidth,angle=0,clip]{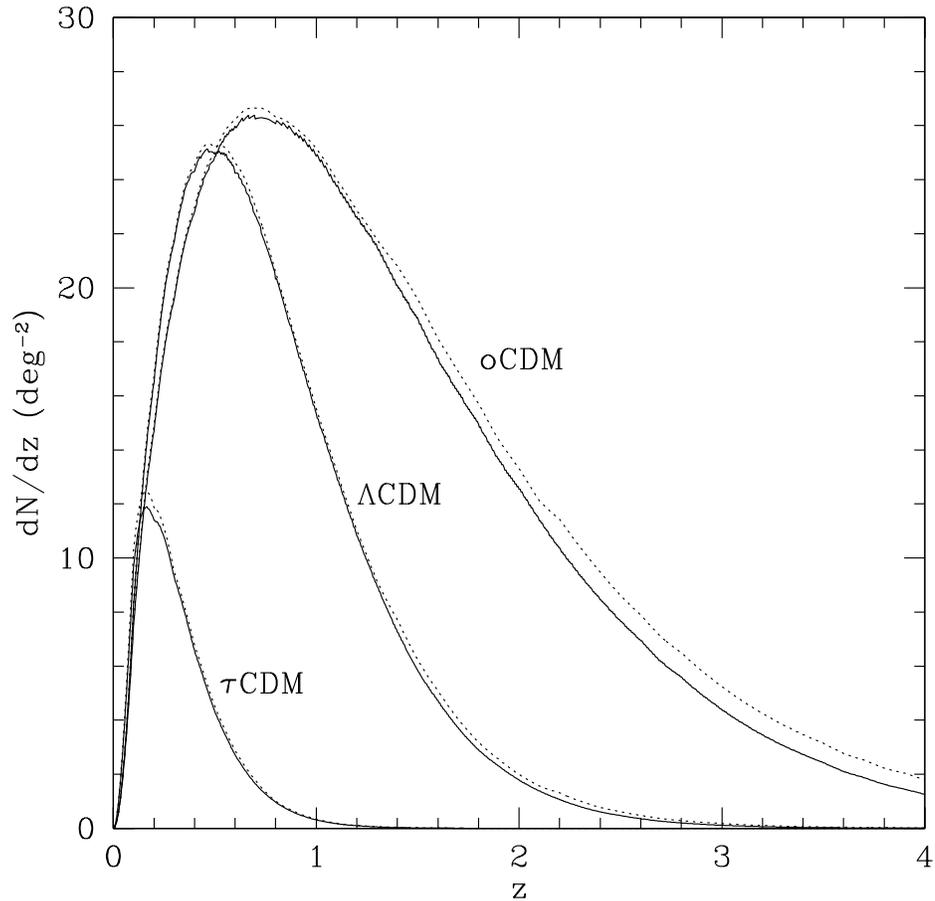}
\vskip 0pt \caption{
   Expected number counts of galaxy clusters from an upcoming
   dedicated interferometric SZE survey array.  The curves are very
   different for the three different cosmologies, having been
   normalized to the local cluster abundance.  The two sets of curves
   are slightly different treatments of the mass limits.  Notice that
   for low-$\OmM$ cosmologies, there are a significant number of
   galaxy clusters with $z>1$.  
   The cosmologies in the figure are ($\OmM$, $\OmL$, $\sigma_8$):
   $\Lambda$CDM (0.3, 0.7, 1.0); OCDM (0.3, 0.0, 0.6); and $\tau$CDM
   (1.0, 0.0, 1.0). See \citep{holder2000} for details.
\label{fig:dNdz}}
\end{figure*}

The beauty of the SZE and X-ray technique for measuring distances is
that it is completely independent of other techniques, and that it can
be used to measure distances at high redshifts directly.  Since the
method relies on the well-understood physics of highly ionized
plasmas, it should be largely independent of cluster evolution.
Inspection of Figure~\ref{fig:da} already provides confidence that a
large survey of SZE distances consisting of perhaps a few hundred
clusters with redshifts extending to one and beyond would allow the
technique to be used to trace the expansion history of the Universe,
providing a valuable independent check of the recent determinations of
the geometry of the Universe from Type Ia supernovae \citep{riess1998,
perlmutter1999} and CMB primary anisotropy experiments
\citep{stompor2001, netterfield2002, pryke2002, spergel2003}.

\section{SZE Surveys}
\label{sec:surveys}

A promising way of finding high-redshift clusters is to perform deep,
large-scale SZE surveys, taking advantage of the redshift independence
of the SZE.  Such surveys will provide large catalogs of galaxy
clusters, many of which will be at high redshift.  The redshift
evolution of the number density of galaxy clusters is critically
dependent on the underlying cosmology, and in principle can be used to
determine the equation of state of the ``dark energy''
\citep[e.g.,][]{haiman2001, holder2000, holder2001}.
Figure~\ref{fig:dNdz} illustrates the dependence of the evolution of
cluster number density on cosmology \citep{holder2000}.  All three
cosmologies are normalized to the local cluster abundance.  Notice
that for low-$\OmM$ cosmologies, there are a significant number of
galaxy clusters with $z>1$.

There are a number of dedicated SZE experiments under construction
that will perform deep, large-scale SZE surveys.  In the next few
years, there are a number of interferometric approaches that should
find hundreds of galaxy clusters.  Bolometer array based SZE experiments
should find roughly thousands of clusters just one year after the
interferometric experiments.  The following generation of bolometer
array based SZE experiments should measure tens of thousands of galaxy clusters
on a roughly five-year time scale.  The possible systematics that could
affect the yields of SZE surveys are presently too large to realize
the full potential of a deep SZE survey covering thousands of square
degrees.  These systematics include, for example, the uncertainties on
the survey mass detection limit owing to unknown cluster structure and
cluster gas evolution.  These systematics can begin to be addressed
through detailed follow-up observations of a moderate-area SZE survey.
High-resolution SZE, X-ray, and weak lensing observations will provide
insights into the evolution and structure of the cluster gas.  Numerical
simulations directly compared and normalized to the SZE yields should
provide the necessary improvement in our understanding of the mass
function.

\section{Summary}
\label{sec:summary}

Soon after it was proposed, it was realized that SZE observations
could yield distances to galaxy clusters when combined with X-ray
data.  SZE/X-ray distances are independent of the extragalactic
distance ladder and do not rely on clusters being standard candles or
rulers.  The promise of direct distances helped motivate searches for
the SZE in galaxy clusters.  It is only in recent years that
observations of the small ($\lsim 1$ mK) SZE signal have yielded
reliable detections and imaging, mostly due to advances in
observational techniques.

SZE/X-ray distances have progressed from distance determinations one
cluster at a time to samples of tens of galaxy clusters.  To date,
there are 41 SZE/X-ray determined distances to 26 different galaxy
clusters.  The combined 41 cluster distances imply a Hubble parameter
of $\sim 61 \pm 3 \pm 18\ \ksM$, where the approximate
uncertainties are statistical followed by systematic at 68\%
confidence.  Shared systematics between the determined distances make
it difficult to determine them precisely.  For example, multiple distance
determinations to a single galaxy cluster overstress any asphericity
and orientation effects.  There are three homogeneously analyzed
samples of clusters consisting of five, seven, and 18 galaxy clusters.  Even
with these small samples, systematics clearly dominate.  Systematic
uncertainties are approachable with current X-ray and radio
observatories.

SZE/X-ray determined direct distances to galaxy clusters are beginning
to trace out the theoretical angular-diameter distance relation.  It
is clear that with a large sample of galaxy clusters at high redshift,
SZE/X-ray distances will be able to determine the geometry of the
Universe.  In fact, if for some (highly unlikely) reason there was a
systematic offset with the SZE/X-ray distances, the shape of the
curve could still be determined and cosmological parameters inferred.  

The redshift independence of the SZE will be exploited with surveys of
large patches of sky with next generation, dedicated SZE experiments.
Such surveys will provide large catalogs of clusters, many of which
will be at high redshift ($z>1$), providing large enough samples
to use SZE/X-ray distances to determine both the Hubble constant and
the geometry of the Universe.  In addition, the cluster yields from
deep, large-scale SZE surveys are critically dependent on the
cosmology, potentially allowing a determination of the equation of
state of the ``dark energy.''  High-resolution SZE, X-ray, and optical
(including weak lensing) observations will provide insights into the 
evolution and structure of the cluster gas necessary to exploit fully
the evolution in cluster yields from SZE surveys.

\vspace{0.3cm} 
{\bf Acknowledgements}.  
The author is grateful for financial support from NASA Chandra
Postdoctoral Fellowship PF 1-20020.

\end{document}